# Structural and Magnetic Properties of Pyrochlore Solid Solutions $(Y,Lu)_2Ti_{2-x}(Nb,Ta)_xO_{7\pm y}$

D.V. West[a,*], T.M. McQueen[a], Q. Huang[b], R.J. Cava[a]

[a]Department of Chemistry, Princeton University NJ 08544, USA
[b]NIST Center for Neutron Research, Gaithersburg MD 20899, USA

[*]Corresponding author. Email address: dwest@princeton.edu. Fax: 609-258-6746, Address: Frick Laboratory, Princeton University, 08544

**Abstract**

The synthesis and characterization of the pyrochlore solid solutions, $Y_2Ti_{2-x}Nb_xO_{7-y}$, $Lu_2Ti_{2-x}Nb_xO_{7-y}$, $Y_2Ti_{2-x}Ta_xO_{7-y}$ and $Lu_2TiTaO_{7-y}$ (-0.4<y<0.5), is described. Synthesis at 1600 ºC, and $10^{-5}$ Torr yields oxygen deficiency in all systems. All compounds are found to be paramagnetic and semiconducting, with the size of the local moments being less, in some cases substantially less, than the expected value for the number of nominally unpaired electrons present. Thermogravimetric analysis (TGA) shows that all compounds can be fully oxidized while retaining the pyrochlore structure, yielding oxygen rich pyrochlores as white powders. Powder neutron diffraction of $Y_2TiNbO_7$–based samples was done. Refinement of the data for oxygen deficient $Y_2TiNbO_{6.76}$ indicates the presence of a distribution of oxygen over the 8b and 48f sites. Refinement of the data for oxygen rich $Y_2TiNbO_{7.5}$ shows these sites to be completely filled, with an additional half filling of the 8a site. The magnetic and TGA data strongly suggest a preference for a $Ti^{3+}/(Nb,Ta)^{5+}$ combination, as opposed to $Ti^{4+}/(Nb,Ta)^{4+}$, in this pyrochlore family. In addition, the evidence clearly points to $Ti^{3+}$ as the source of the localized moments, with no evidence for localized $Nb^{4+}$ moments.

Introduction

Interest in niobium–based ceramics has stemmed primarily from the ferroelectric properties of $Nb^{5+}$, the fully oxidized cation, with less focus on reduced niobates having oxidation states less than 5+. Phases containing $Nb^{4+}$ are often derivatives of the perovskite and pyrochlore structures [1], while several other, more reduced phases, possess $Nb_6O_{12}$ clusters in which the Nb ions are found on the vertices of an octahedron [2,3]. As 4$d$ valence electrons have a tendency to form Nb–Nb bonds, many reduced niobates are metals or small-band gap semiconductors. There are some claims of temperature independent paramagnetism and extremely few examples exhibiting localized magnetic moments, the most prominent of which are the block–based niobium oxide shear structures, $e.g.$ $Nb_{12}O_{29}$ and $Nb_{22}O_{54}$ [4-6]. There have also been reports of magnetic niobate pyrochlores [7].

As a structural class, pyrochlores (general formula $A_2B_2O_6O'$) hold interest as geometrically frustrating lattices, sometimes allowing the observation of exotic phenomena, such as spin ice. A unique structure type, it has two interpenetrating frustrating A and B sub–lattices, but with different crystal field environments. Magnetic B–site materials such as $Y_2(Mn,Mo)_2O_7$ [8,9] have yielded insulating examples of both frustrated ferro– and anti–ferromagnetic materials. Meanwhile, the A–site materials $(Dy,Ho)_2Ti_2O_7$ have received much attention as examples of spin–ice compounds [10,11].

Recent studies on niobium based pyrochlores, like other reduced niobates, have yielded weakly magnetic, or non–magnetic compounds. $CaLnNb_2O_7$ ($Ln$ = Y, La, Lu) [12,13] is non–magnetic and electrically insulating, possibly explained by strong spin–orbit coupling, or electronic Nb–Nb coupling. The latter interpretation is supported by DFT calculations on $Y_2Nb_2O_7$ [14]. However, the $Ln_2Nb_2O_7$ ($Ln$ = Y, La, Lu) phases based on $Nb^{4+}$ cannot be synthesized under equilibrium conditions up to 1625 ºC [15].

In this study, the structural and electronic properties of the solid solutions $Ln_2Ti_{2-x}M_xO_{7\pm y}$, where $Ln$=Y,Lu and M=Nb,Ta, are explored in order to gain a clearer understanding of the electronic and structural properties of these systems, and in particular, the magnetic attributes of niobium 4$d$ electrons in the pyrochlore setting.

Experimental

$Y_2Ti_{2-x}Nb_xO_{7-y}$, $Lu_2Ti_{2-x}Nb_xO_{7-y}$, $Y_2Ti_{2-x}Ta_xO_{7-y}$, with x = 0, 0.25, 0.5, 0.75, and 1, and $Lu_2TiTaO_{7-y}$, were made by mixing stoichiometric quantities of $Y_2O_3$ (Alfa Aesar, 99.999%), $Lu_2O_3$ (Alfa Aesar 99.9%), $NbO_2$ (Alfa Aesar 99+%), $Nb_2O_5$ (Aldrich 99.5%), $Ta_2O_5$ (Alfa Aesar 99.993%), and $TiO_2$ (Alfa Aesar 99.9%). They were then ground to homogeneity and wrapped in Nb foil. Unless otherwise specified, the samples were heated twice at 1600 ºC for 12 hours in a vacuum furnace under a dynamic vacuum of $10^{-5}$ Torr with intermittent grinding. $Y_2Ti_2O_7$ was prepared at 1500 ºC in air, followed by annealing at 1500 ºC under vacuum. Powder samples, as opposed to pellets, were preferred for reproducible oxygen loss caused by the high vacuum. Oxygen rich pyrochlores were produced by annealing the oxygen deficient powders at either 280 or 500 ºC under flowing $O_2$. $CaYNb_2O_7$ was prepared as a control by mixing $Ca_4Nb_2O_9$ (prepared in air at 1300° C from $CaCO_3$ (Mallinckrodt, 99.9%) and $Nb_2O_5$), $Nb_2O_5$, $NbO_2$ and $Y_2O_3$ stoichiometrically. The powder was pressed into a pellet, wrapped in Mo foil, and heated in a vacuum furnace back-filled with argon at 1525 °C for 32 hours, with an intermediate grinding, re-pelleting and then re-heating at 1625 °C for 12 hours.

Magnetic susceptibilities were measured using a Quantum Design PPMS magnetometer under a 1 T field in the temperature range 5–60 K. Susceptibility data were fit to the Curie–Weiss law in order to obtain the $p_{eff}$ values of the samples.

Oxygen stoichiometries were determined using a TA Instruments Thermogravimetric Analyzer (TGA). Samples sizes of 20–45 mg were placed on a platinum weighing pan and heated from room temperature to 550 ºC or 700 ºC at either 0.25 or 2 ºC/min under flowing $O_2$. Under these conditions all mass change was due to oxygen absorption, and the final mass for the white samples obtained corresponded to a state of maximal oxidation. This allowed the determination of the initial oxygen stoichiometry of the samples synthesized under high vacuum.

The $Ti^{3+/4+}$ and $Nb^{4+/5+}$ or $Ta^{4+/5+}$ ratios for all samples were estimated from the TGA curves. In all but one of the Nb-based samples, and in the Ta-based samples where x = 1.0 and 0.75, an intermediate region in between two mass gain transitions was observed. For these samples, linear fits to portions of the first derivatives were used to demarcate the extrema of the intermediate region. Assuming that complete oxidation of $Ti^{3+}$ occurs before any oxidation of $Nb^{4+}$ or $Ta^{4+}$, the oxygen content at which $Ti^{3+}$ oxidation ends and $Nb^{4+}$ or $Ta^{4+}$ oxidation begins was approximated as half of the difference between the oxygen contents at the extrema, thus allowing an estimation of the cation oxidation state distributions in the reduced materials. For $Y_2Ti_{1.75}Nb_{0.25}O_{6.85}$ and the other Ta-based samples, no intermediate region was observed and by comparison to the other TGA plots and their trends, it is reasonable to assume that all Nb or Ta is 5+ in these samples.

Crystal structures were characterized using both X–ray and neutron diffraction of polycrystalline samples at room temperature. Powder X–ray diffraction (PXRD) data were collected with a Bruker D8–Focus, using Cu Kα radiation with a graphite diffracted beam monochromator. Powder neutron diffraction (PND) data were collected on $Y_2TiNbO_{6.76}$ and $Y_2TiNbO_{7.5}$ at the NIST Center for Neutron Research on the high resolution powder neutron diffractometer (BT–1) with neutrons of wavelength 1.5403 Å produced by using a Cu(311) monochromator. Collimators with horizontal divergences of 15′, 20′ and 7′ of arc were used before and after the monochromator and after the sample, respectively. Data were collected in the 2θ range of 3–168° with a step size of 0.05°. The structural parameters were determined by Rietveld refinement of the neutron diffraction data using the GSAS program [16,17]. The atomic neutron scattering factors used in the refinements were Y – 0.775, Ti – (-0.337), Nb – 0.705, and O – 0.581 x $10^{-12}$ cm.

Results and Discussion

One known and three new single phase pyrochlore solid solutions of the general formula $Ln_2Ti_{2-x}M_xO_{7+y}$ (*Ln*=Y,Lu; M=Nb,Ta; $0 \leq x \leq 1$; $-0.4 \leq y \leq 0.5$) were synthesized under high vacuum as described. Under these conditions, significant loss of oxygen occurs, yielding black powders containing a mix of $Ti^{3+/4+}$ and either $Nb^{4+/5+}$ or $Ta^{4+/5+}$. For all samples, a 500 ºC anneal under flowing $O_2$ turns the powders white, oxidizing all of the reduced cations to $Ti^{4+}$, $Nb^{5+}$ and $Ta^{5+}$. All of the powder diffraction patterns (Figure 1) show the same pyrochlore peaks consistent with Fd-3m symmetry. Increased doping on the Ti site with Nb or Ta systematically increases the unit cell parameter. Interestingly, the unit cell parameter increases with excess oxygen when M=Nb, but decreases when M=Ta. Since the unit cell parameter is a competition between decreasing cation size with increasing oxidation state, disruption of metal-metal bonds (if present) and extra oxygen atoms trying to occupy the same space, it is not possible to unambiguously identify the origin of the difference between the Nb and Ta cases.

The compounds $Y_2Ti_{2-x}Nb_xO_{7-y}$, $Lu_2Ti_{2-x}Nb_xO_{7-y}$, $Y_2Ti_{2-x}Ta_xO_{7-y}$ and $CaYNb_2O_7$ were measured both for their magnetic properties and their mass gain in the TGA under flowing oxygen. The measured mass gain was dependent upon the number of heatings under high vacuum with intermittent grinding, but was independent of the duration of these heatings beyond 12 hours.

The TGA data for the control samples containing only Ti ($Y_2Ti_2O_{6.88}$ and $Lu_2Ti_2O_{6.64}$) or only Nb ($CaYNb_2O_7$) are shown in Figure 2a. The titanate samples show only one mass gain around 200 ºC and both show sizable effective magnetic moments per formula unit ($p_{eff}$/f.u.) (0.29 ±0.01 and 0.57 ±0.01 $\mu_B$ respectively). The niobate also shows only one transition, but at 400 ºC, and the compound is non–magnetic.

$Y_2TiNbO_{7-y}$ was made by heating the sample four times under high vacuum with intermittent grindings. After each heating, some of the powder was measured for oxygen gain in the TGA on heating in $O_2$ (Figure 2b). In all cases, the curves show the same general shape: two mass increases, one just above 200 °C and another just above 400 °C, surrounding an intermediate region with a substantially smaller slope. The control samples indicate that the low temperature mass increase is associated with oxidation of $Ti^{3+}$, and the high temperature increase with oxidation of $Nb^{4+}$. The exact nature of the intermediate region, whose slope changes gradually with increased amounts of heating, is unknown. Some of the sample from the first heating was annealed overnight under flowing $O_2$ gas at 280 ºC. The annealed sample was found to be non–magnetic. Subsequent analysis in the TGA shows that the annealed sample has one mass gain just after 400 ºC, equal in magnitude to the second transition seen in its parent sample. This indicates that all of the magnetic electrons are associated with the low temperature oxidation, thus pointing to $Ti^{3+}$, and not $Nb^{4+}$, as the source of localized magnetism. Two of the other samples, the ones heated twice and four times with intermediate grindings, were measured in the magnetometer as well. They were found to have the same $p_{eff}$/f.u. within error (0.33 ±0.01 $\mu_B$), in spite of having significantly different oxygen contents.

For the series $Y_2Ti_{2-x}Nb_xO_{7-y}$, as x decreases fewer oxygen vacancies are created by the high vacuum heatings (Figure 3a), with the full titanate sample having the formula $Y_2Ti_2O_{6.88}$. As x decreases, the temperature range of the intermediate region grows smaller, disappearing entirely by the x = 0.25 sample. These TGA data show that the oxidation states of the B–site cations exist as a mixture of $Ti^{3+/4+}$ and $Nb^{4+/5+}$, with the oxidation of the reduced cations happening at separate temperatures in spite of being randomly mixed. The magnetic data for this family is shown in Figure 4a. The size of $p_{eff}$/f.u. for all of these samples was small and changed little across the series (Figure 5).

The $Lu_2Ti_{2-x}Nb_xO_{7-y}$ series (Figure 3b) contrasts to the Y analogue in several ways. In this series, the temperature range of the intermediate region shrinks as before, but is clearly seen in all niobium-containing samples. And while the titanate end–member retains a single mass gain at 200 ºC, the oxygen content is significantly lower than its Y counterpart at $Lu_2Ti_2O_{6.64}$, possibly due to the smaller size of $Lu^{3+}$, which is expected to allow it to accommodate fewer oxygen neighbors. Like its Y analogue, the B–site cations in these compounds possess a $Ti^{3+/4+}$–$Nb^{4+/5+}$ mixture of oxidation states. Across the series, the magnetic moment was its lowest at x = 1.0 ($p_{eff}$/f.u. = 0.28 ±0.01 $\mu_B$), about the same at x = 0.75, and then rose steadily to its highest moment at x = 0 (0.57 ±0.01 $\mu_B$), significantly larger than the Y analogue (see Figures 4 and 5).

$Y_2Ti_{2-x}Ta_xO_{7-y}$ shows markedly different behavior (Figure 3c) reflecting the influence of the more electropositive Ta in place of Nb. Examining first the x = 1.0 compound, the TGA curve corresponds to an initial stoichiometry of $Y_2TiTaO_{6.9}$, implying a 20% reduction of the $Ta^{5+}$ to $Ta^{4+}$. The same initial mass gain at 200 °C is seen followed by a very gradual and small mass gain up to 600 °C, presumed to be $Ta^{4+}$ oxidation. Because it occurs at higher temperature than the initial mass gain, the oxidation of $Ti^{2+}$ to $Ti^{3+}$ in the parent material is an unlikely explanation of the second mass gain because $Ti^{2+}$ oxidation would have to occur before the $Ti^{3+}$ oxidation. The $x = 0.75$ sample retains a small amount of this high temperature gradual mass gain, but for $x < 0.75$, only the low temperature transition is observed. The magnetic susceptibility data are shown in Figure 4c. Across the series, the $p_{eff}$/f.u was its highest at $x = 1.0$ (0.54 $\mu_B$), about the same at $x = 0.75$, and decreased steadily to its lowest at $x = 0$ (0.26 $\mu_B$) (Figure 5). Tantalum has never been shown to exhibit localized magnetism, and as such these data clearly indicate $Ti^{3+}$ as the source of the local moments for this series.

Comparing all three series, it is useful to plot the $p_{eff}$/f.u. vs. $Ti^{3+}$/f.u or $Nb^{4+}$/f.u. (Table I, Figure 6). The correlations of the $p_{eff}$/f.u. to the $Ti^{3+}$/$Nb^{4+}$ contents were quite different between the three series. For the Nb–based compounds, it is very difficult to draw any correlation between $Ti^{3+}$ content and $p_{eff}$. For the series $Y_2Ti_{2-x}Nb_xO_{7-y}$, the local moment changes little even as the $Ti^{3+}$ content varies significantly. Conversely, the Lu analogue has fairly constant $Ti^{3+}$ content, but displays a changing local moment. Looking instead at $Nb^{4+}$ content, there is no discernible trend in the Y–based series, but for the Lu–based series the local moment decreases as $Nb^{4+}$ content increases (Figure 6a). Compared to the Nb–based compounds, $Y_2Ti_{2-x}Ta_xO_{7-y}$ exhibits a more rational trend (Figure 6b), with the local moment increasing steadily with increasing $Ti^{3+}$ content. Also, the Ti–only compound $Lu_2Ti_2O_{6.64}$ sits close to the trend formed by the $Y_2Ti_{2-x}Ta_xO_{7-y}$ series in the $p_{eff}$ vs. $Ti^{3+}$ plot (Figure 6b), further stressing the relationship of the magnetism to localized $Ti^{3+}$ states.

A DFT study on $Y_2Nb_2O_7$ by Blaha et al. [14], supported by structural studies done in our lab on Nb–based pyrochlores (including $Y_2TiNbO_7$), indicate that a B–site displacement, in which $Nb^{4+}$ atoms move off the 16c sites toward the center of the B-sublattice tetrahedra forming a 4–center 2–electron bond, is prevalent in $Nb^{4+}$ pyrochlores [15]. Though not conclusive, the data presented here are consistent with these observations, and suggest that niobium may be suppressing magnetism by replacing the paramagnetic states with singlet states, while the tantalum is a spectator ion.

$Y_2TiNbO_{6.76}$ and $Y_2TiNbO_{7.5}$ were studied by powder neutron diffraction (Figure 7). The data were refined using two models: the standard pyrochlore model, and a B–site displacement model (Table II). We refined this second model to compare these samples to the DFT calculations and the other work in our lab [15]. The B cations occupy the vertices of perfect tetrahedra that share corners infinitely in all directions. As each B atom is shared by two tetrahedra, the displacement moves them closer to the center of one tetrahedron, and farther away from the other. In both cases, it was found that the temperature factors for the B–site cations and $\chi^2$ values were reduced for the displaced atom model. In both cases the displacement was statistically significant, though slightly smaller in the oxygen rich case. This refinement leaves open the possibility that the non-systematic magnetic trends in the Nb-based compounds are explained by the formation of singlet states.

For the oxygen deficient sample, the oxygen occupancies were refined while constrained to sum to the value determined by TGA. For the oxygen rich sample, the excess oxygen was placed on the 8a site (pyrochlores are defect fluorite structures, which have one out of eight anion sites vacant; this vacancy corresponds to the 8a site, and is the only possible site for excess oxygen to go). While constraining them to sum to the oxygen content corresponding to the state of maximal oxidation, refining the oxygen occupancies resulted in greater than unit occupancies on either the 8b or the 48f sites, and less than half occupancy on the 8a site. As there is no other rational site for the oxygens in the pyrochlore structure, the 8b and 48f oxygen occupancies were constrained to one, with half occupancy on the 8a site. We note that there is no thermodynamically stable phase $Y_2TiNbO_{7.5}$. The equilibrium assemblage for samples synthesized at 1600 ºC with this formula is $Y_2Ti_2O_7$ and $YNbO_4$.

Conclusion

One known, and three new pyrochlore solid solutions, have been synthesized and characterized. Powder neutron diffraction studies show a distribution of the oxygen between the 8b and 48f sites for $Y_2TiNbO_{6.76}$, and a half occupancy of the 8a site for $Y_2TiNbO_{7.5}$. In addition, the data for this Nb–based compound are consistent with both a standard pyrochlore model, and a B–site displacement model in which the B–site cations displace off of their special positions.

TGA and magnetic studies suggest some preference for a $Ti^{3+}/M^{5+}$ combination in these solid solutions, rather than a purely $Ti^{4+}/M^{4+}$ combination on the B–site. Moreover, the loss of magnetism after the low temperature oxidation, shown to be $Ti^{3+}$ oxidation, indicates $Ti^{3+}$ as the source of local moments in these compounds. This conclusion is further supported by the increase in the $p_{eff}$/f.u. with the increase in $Ti^{3+}$ content for the Ti–only and Ti/Ta–based compounds. These results indicate that the pyrochlore family of niobates, unlike the crystallographic shear structures, does not support the existence of local moment magnetism for $Nb^{4+}$.

Acknowledgements

This research was supported by the NSF program in Soild State Chemistry, grant number NSF DMR-0703095. Certain commercial materials and equipment are identified in this report to describe the subject adequately. Such identification does not imply recommendation or endorsement by the NIST, nor does it imply that the materials and equipment identified are necessarily the best available for the purpose. T. M. McQueen gratefully acknowledges support of the National Science Foundation Graduate Research Fellowship Program.
References

[1] O.G. D'yachenko, S.Y. Istomin, A.M. Abakumov, E.V. Antipov. *Inorg. Mater.* 36 (2000) 247-259.

[2] B. Hessen, S.A. Sunshine, T. Siegrist, A.T. Fiory, J.V. Waszczak. *Chem. Mater.* 3 (1991) 528-534.

[3] J. Kohler, G. Svensson, A. Simon. *Angew. Chem., Int. Ed.* 31 (1992) 1437-1456.

[4] R.J. Cava, B. Batlogg, J.J. Krajewski, P. Gammel, H.F. Poulsen, W.F. Peck, L.W. Rupp. *Nature* 350 (1991) 598-600.

[5] J.E.L. Waldron, M.A. Green, D.A. Neumann. *J. Am. Chem. Soc.* 123 (2001) 5833-5834.

[6] T. McQueen, Q. Xu, E. Andersen, H.W. Zandbergen, R.J. Cava. *J. Solid State Chem.* 180 (2007) 2864-2870.

[7] O. Sakai, Y. Jana, R. Higashinaka, H. Fukazawa, S. Nakatsuji, Y. Maeno. *J. Phys. Soc. Jpn.* 73 (2004) 2829-2833.

[8] J.E. Greedan, M. Sato, Y. Xu, F.S. Razavi. *Solid State Commun.* 59 (1986) 895-897.

[9] J.N. Reimers, J.E. Greedan, R.K. Kremer, E. Gmelin, M.A. Subramanian. *Phys. Rev. B* 43 (1991) 3387-3394.

[10] M.J. Harris, S.T. Bramwell, D.F. McMorrow, T. Zeiske, K.W. Godfrey. *Phys. Rev. Lett.* 79 (1997) 2554-2557.
11

[11] A.P. Ramirez, A. Hayashi, R.J. Cava, R. Siddharthan, B.S. Shastry. *Nature* 399 (1999) 333-335.

[12] S.Y. Istomin, O.G. D'Yachenko, E.V. Antipov, G. Svensson. *Mater. Res. Bull.* 32 (1997) 421-430.

[13] S.Y. Istomin, O.G. D'Yachenko, E.V. Antipov, G. Svensson, B. Lundqvist. *Mater. Res. Bull.* 33 (1998) 1251-1256.

[14] P. Blaha, D.J. Singh, K. Schwarz. *Phys. Rev. Lett.* 93 (2004) 216403.

[15] T. McQueen, R.J. Cava. *Submitted to Journal of Physics C*.

[16] A.C. Larson, R.B. Von Dreele. *Los Alamos National Laboratory Report LAUR* (2000) 86-748.

[17] B.H. Toby. *J. Appl. Crystallogr.* 34 (2001) 210-213.

Table I: Correlation of the effective moment to the reduced cation concentrations determined by TGA.

|  | $\mu_B$/f.u. | mol $Ti^{3+}$/f.u. | mol $Nb^{4+}$/f.u. |
|---|---|---|---|
| $Y_2Ti_{2-x}Nb_xO_{7-y}$ | | | |
| x=1.00 | 0.33(1) | 0.98 | 0.61 |
| x=0.75 | 0.23(2) | 0.81 | 0.41 |
| x=0.50 | 0.15(3) | 0.66 | 0.22 |
| x=0.25 | 0.24(1) | 0.56 | 0 |
| x=0.00 | 0.27(3) | 0.24 | - |
| | | | |
| $Lu_2Ti_{2-x}Nb_xO_{7-y}$ | | | |
| x=1.00 | 0.28(1) | 0.94 | 0.72 |
| x=0.75 | 0.28(2) | 0.79 | 0.50 |
| x=0.50 | 0.29(1) | 0.81 | 0.42 |
| x=0.25 | 0.400(3) | 0.81 | 0.26 |
| x=0.00 | 0.567(3) | 0.73 | - |
| | | | |
| $Y_2Ti_{2-x}Ta_xO_{7-y}$ | | | mol $Ta^{4+}$/f.u. |
| x=1.00 | 0.54(1) | 0.94 | 0.23 |
| x=0.75 | 0.55(1) | 0.81 | 0.07 |
| x=0.50 | 0.42(1) | 0.62 | 0 |
| x=0.25 | 0.38(1) | 0.49 | 0 |
| x=0.00 | 0.27(3) | 0.24 | - |

Table II: Structural Parameters from Powder Neutron Diffraction Fd-3m, 8a ($\frac{1}{8}$, $\frac{1}{8}$, $\frac{1}{8}$); 8b ($\frac{3}{8}$, $\frac{3}{8}$, $\frac{3}{8}$); 16c (0,0,0); 16d ($\frac{1}{2}$, $\frac{1}{2}$, $\frac{1}{2}$); 32e (x, x, x); 48f (x, $\frac{1}{8}$, $\frac{1}{8}$)

| | | $Y_2TiNbO_{6.76}$ | | $Y_2TiNbO_{7.5}$ | |
| --- | --- | --- | --- | --- | --- |
| | | normal | displaced model | normal | displaced model |
| Y | site | 16d | 16d | 16d | 16d |
| | Occ | 1 | 1 | 1 | 1 |
| | $B_{iso}$ | 0.87(3) | 0.88(3) | 0.86(2) | 0.87(2) |
| Ti/Nb | site | 16c/16c | 32e/32e | 16c/16c | 32e/32e |
| | x | 0 | 0.013(1)/ 0.013(1) | 0 | 0.007(1)/ 0.007(1) |
| | Occ | 0.5/0.5 | 0.25/0.25 | 0.5/0.5 | 0.25/0.25 |
| | $B_{iso}$ | 3.5(2)/3.5(2) | 2.1(3)/2.1(3) | 1.7(1)/1.7(1) | 1.3(2)/1.3(2) |
| O1 | site | 8b | 8b | 8b | 8b |
| | Occ | 0.98(2) | 0.98(2) | 1 | 1 |
| | $B_{iso}$ | 0.69(10) | 0.59(10) | 0.72(4) | 0.70(4) |
| O2 | site | 48f | 48f | 48f | 48f |
| | x | 0.3350(1) | 0.3347(2) | 0.3354(1) | 0.3353(1) |
| | Occ | 0.966(3) | 0.964(3) | 1 | 1 |
| | $B_{11}$ | 1.40(6) | 1.47(6) | 2.91(5) | 2.96(5) |
| | $B_{22}=B_{33}$ | 0.78(3) | 0.82(4) | 0.82(2) | 0.85(2) |
| O3 | site | - | - | 8a | 8a |
| | Occ | - | - | 0.5 | 0.5 |
| | $B_{iso}$ | - | - | 2.5(1) | 2.4(1) |
| a (Å) | | 10.1805(1) | - | 10.2015(1) | - |
| $\chi^2$ | | 0.9836 | 0.9759 | 1.051 | 1.045 |
| $R_p$ | | 7.02 % | 6.98 % | 5.91 % | 5.88 % |
| $R_{wp}$ | | 8.99 % | 8.96 % | 7.38 % | 7.37 % |

Figure Captions

**Figure 1.** XRD patterns of the x=1 samples from each of the four solid solutions in order from smallest (top) to largest (bottom) unit cell parameters. Pyrochlore superstructure peaks are indicated by the arrows, with corresponding Miller indices.

**Figure 2. (A)** TGA data for $Y_2TiNbO_{7-y}$ showing that with repeated heating, the second transition, corresponding to $Nb^{4+}$ oxidation, grows larger while the first transition, corresponding to $Ti^{+3}$ oxidation, remains unchanged. **(b)** Comparison of the oxidation of $Y_2TiNbO_{7-y}$ to Ti–only compounds $((Y,Lu)_2Ti_2O_{7-y})$ and an Nb–only compound $(CaYNb_2O_{7-y})$ showing the first and second transitions to be the oxidation of $Ti^{+3}$ and $Nb^{4+}$ respectively.

**Figure 3.** TGA data for **(a)** $Y_2Ti_{2-x}Nb_xO_{7-y}$ **(b)** $Lu_2Ti_{2-x}Nb_xO_{7-y}$ **(c)** $Y_2Ti_{2-x}Ta_xO_{7-y}$. A comparison of the two niobium compounds clearly shows the second transition gets smaller with decreasing Nb content. By contrast, the Ta compounds show very little in the second transition indicating the significant preference for $Ta^{5+}$ over $Ta^{4+}$.

**Figure 4.** $1/\chi$ v. T plots for **(a)** $Y_2Ti_{2-x}Nb_xO_{7-y}$, **(b)** $Lu_2Ti_{2-x}Nb_xO_{7-y}$ and **(c)** $Y_2Ti_{2-x}Ta_xO_{7y}$. All samples were heated twice at 1600 ºC under high vacuum. All samples show Curie Weiss behavior down to 5 K, with small Curie constants and negligible values of $\theta_w$.

**Figure 5.** Comparison of the $p_{eff}$/f.u.(effective magnetic moment per formula unit) as a function of x for three of the solid solutions. The comparison of the Y and Lu series based on Ti/Nb reflect the lower tolerance of oxygen deficiency in the Y case, possibly owing to its bigger size. The lines are drawn to guide the eye.

**Figure 6. (a)** Plot showing the negative correlation of $p_{eff}$/f.u. to the amount of $Nb^{4+}$ in the series $Lu_2Ti_{2-x}Nb_xO_{7-y}$, estimated from the TGA data. **(b)** Plot showing the positive correlation of $p_{eff}$/f.u. to the amount of $Ti^{3+}$ for all the compounds measured that do not contain Nb. The lines are drawn to guide the eye.

**Figure 7. (a)** Neutron diffraction data for oxygen deficient $Y_2TiNbO_{7-y}$, refined with Fd-3m symmetry. Refinement of the oxygen occupancies shows a distribution of oxygen on the 8b and 48f sites. Inset: 1/8 of the unit cell showing the distributed 8b and 48f oxygens, and the 8a oxygen sitting in the Y tetrahedron, and the empty Ti/Nb tetrahedron. **(b)** Neutron diffraction data for oxygen rich $Y_2TiNbO_{7-y}$, refined with Fd-3m symmetry and 0.5 occupancy on the 8a site. Inset: 1/8 of the unit cell showing the 1/2 occupied 8a site, sitting within the Ti/Nb tetrahedron.

**Figure 1.**

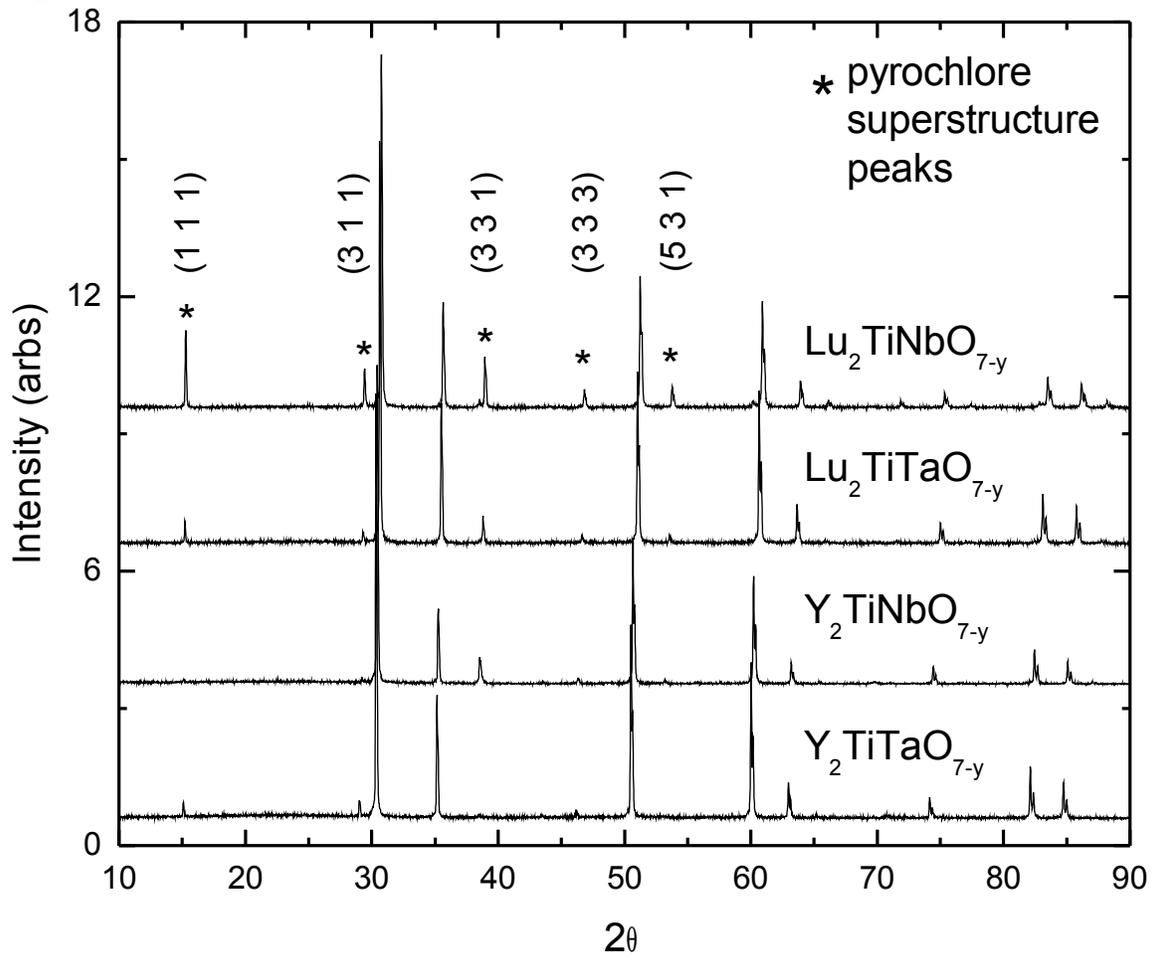

**Figure 2.**

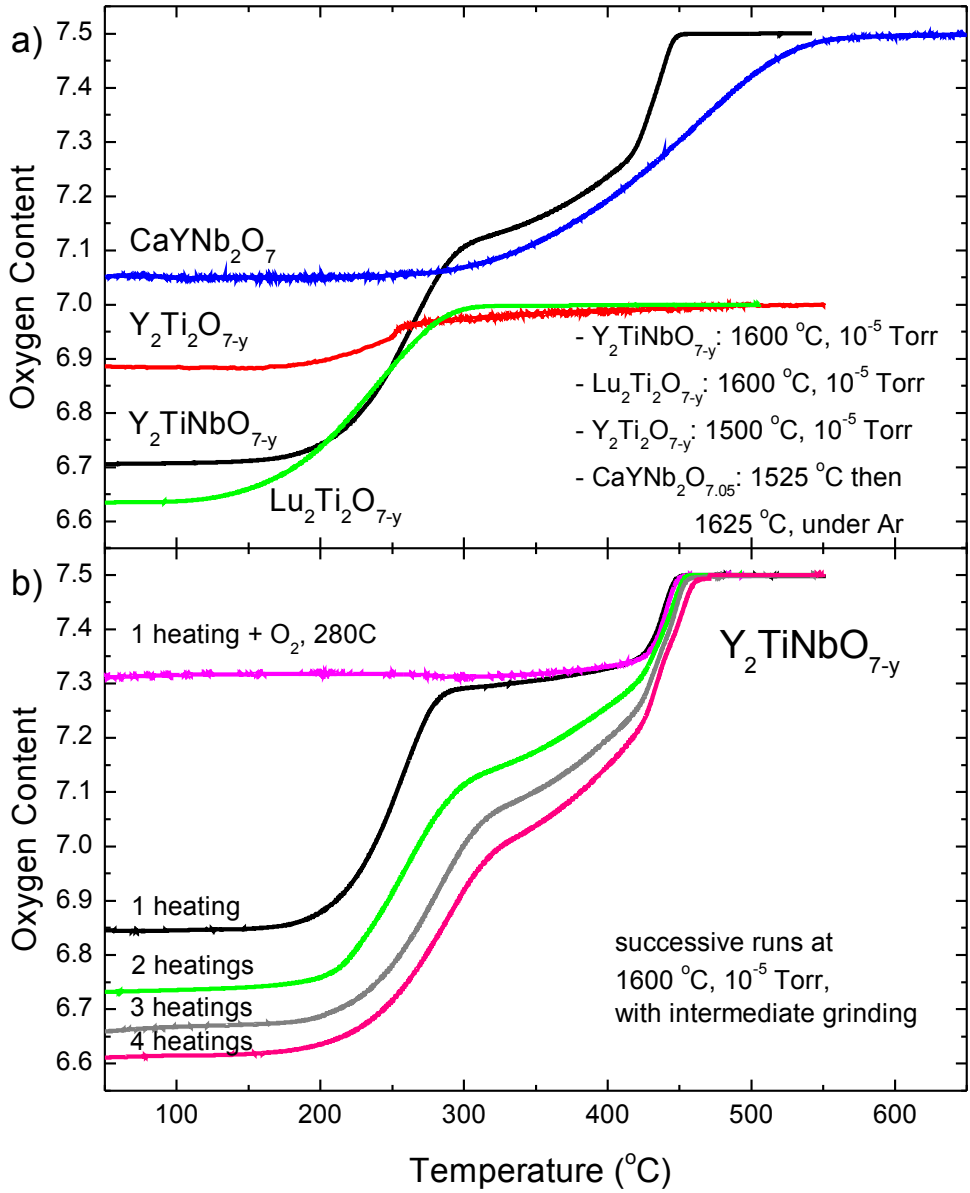

**Figure 3.**

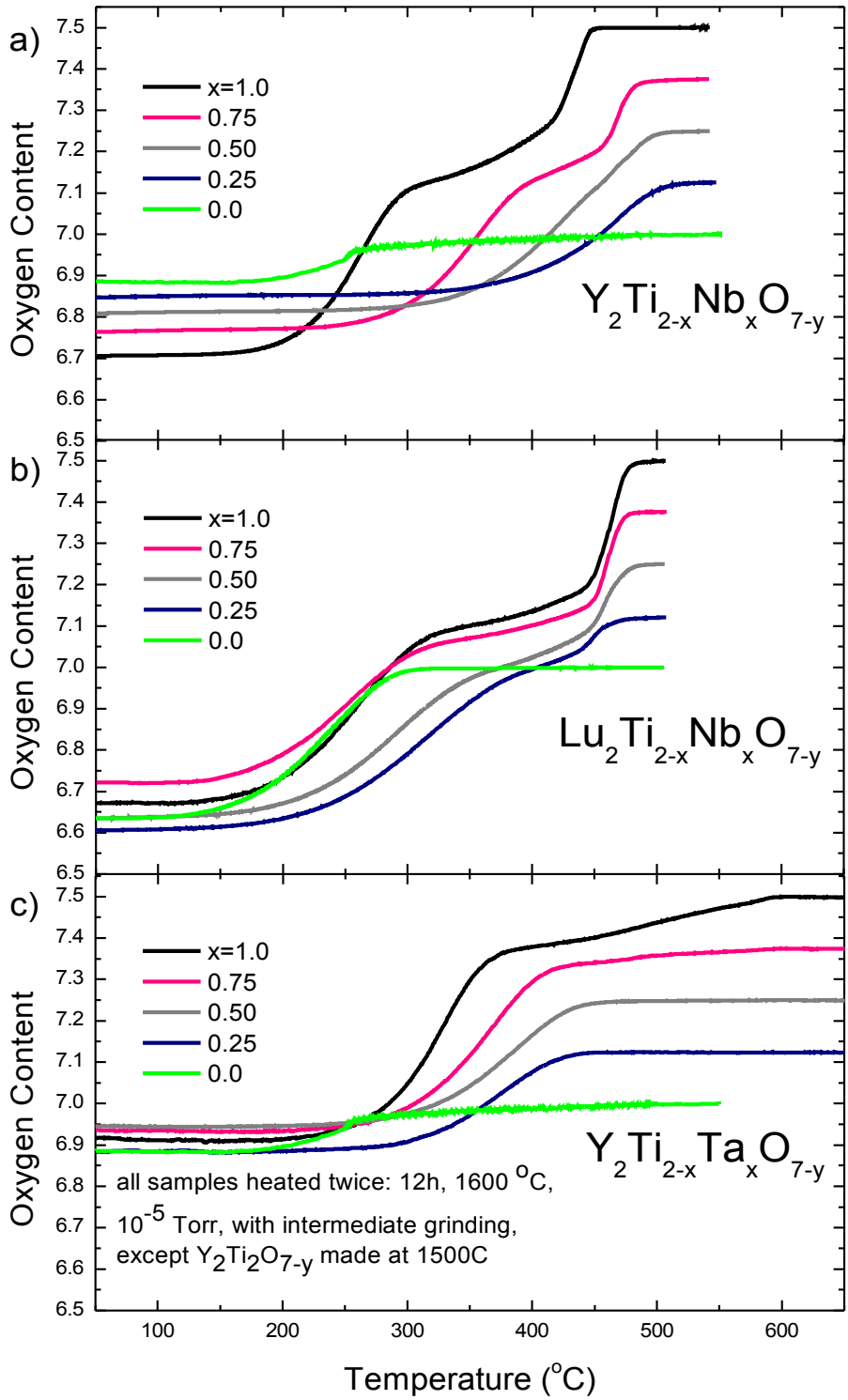

**Figure 4.**

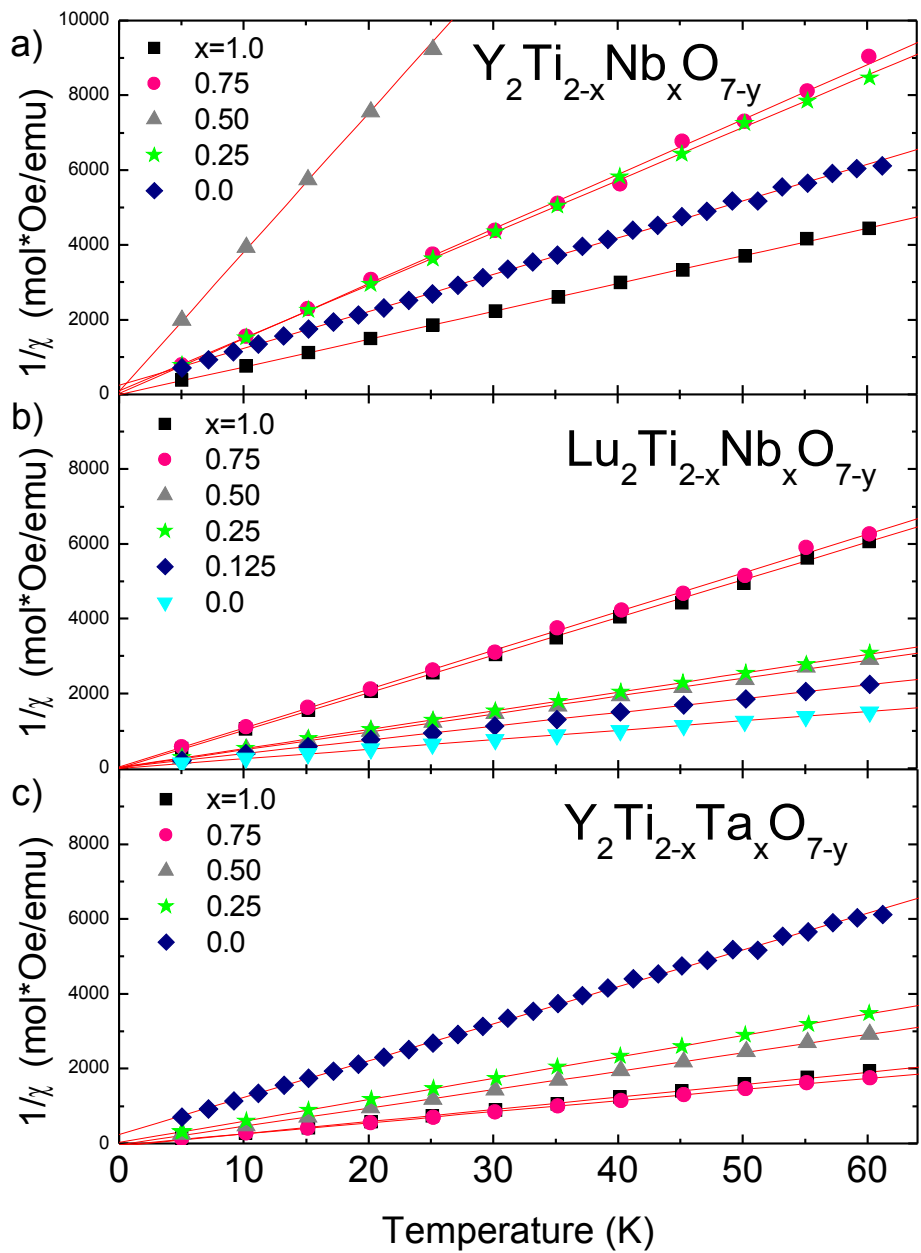

**Figure 5.**

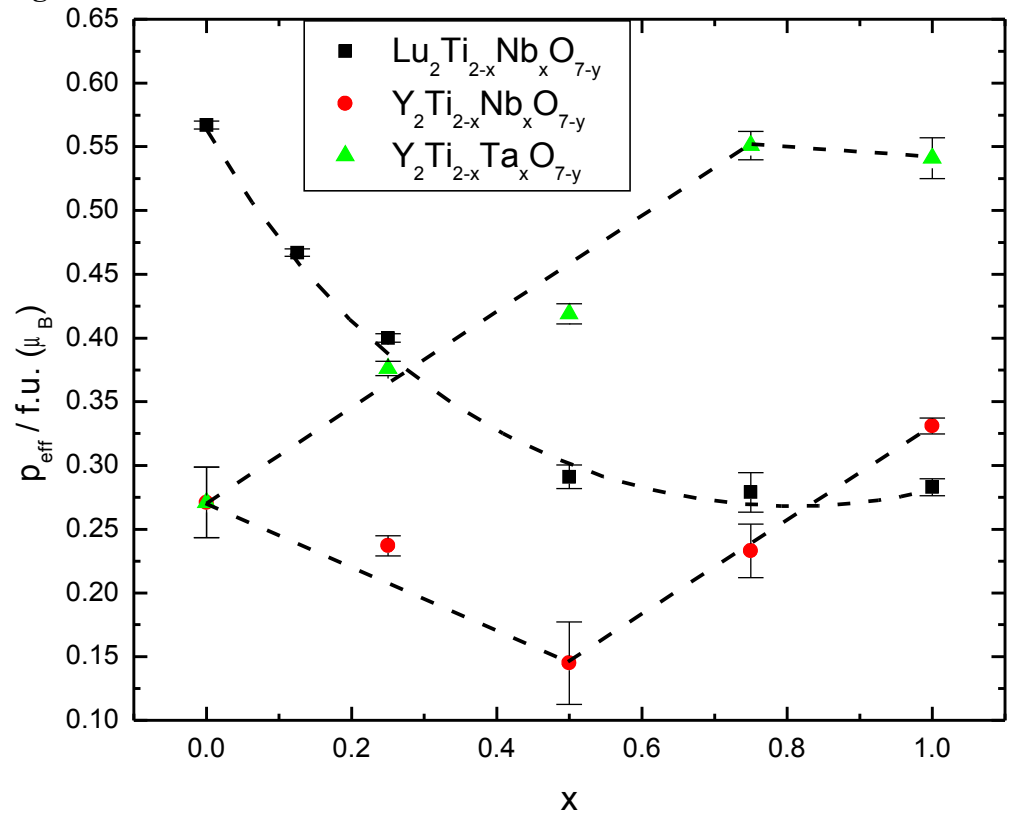

**Figure 6.**

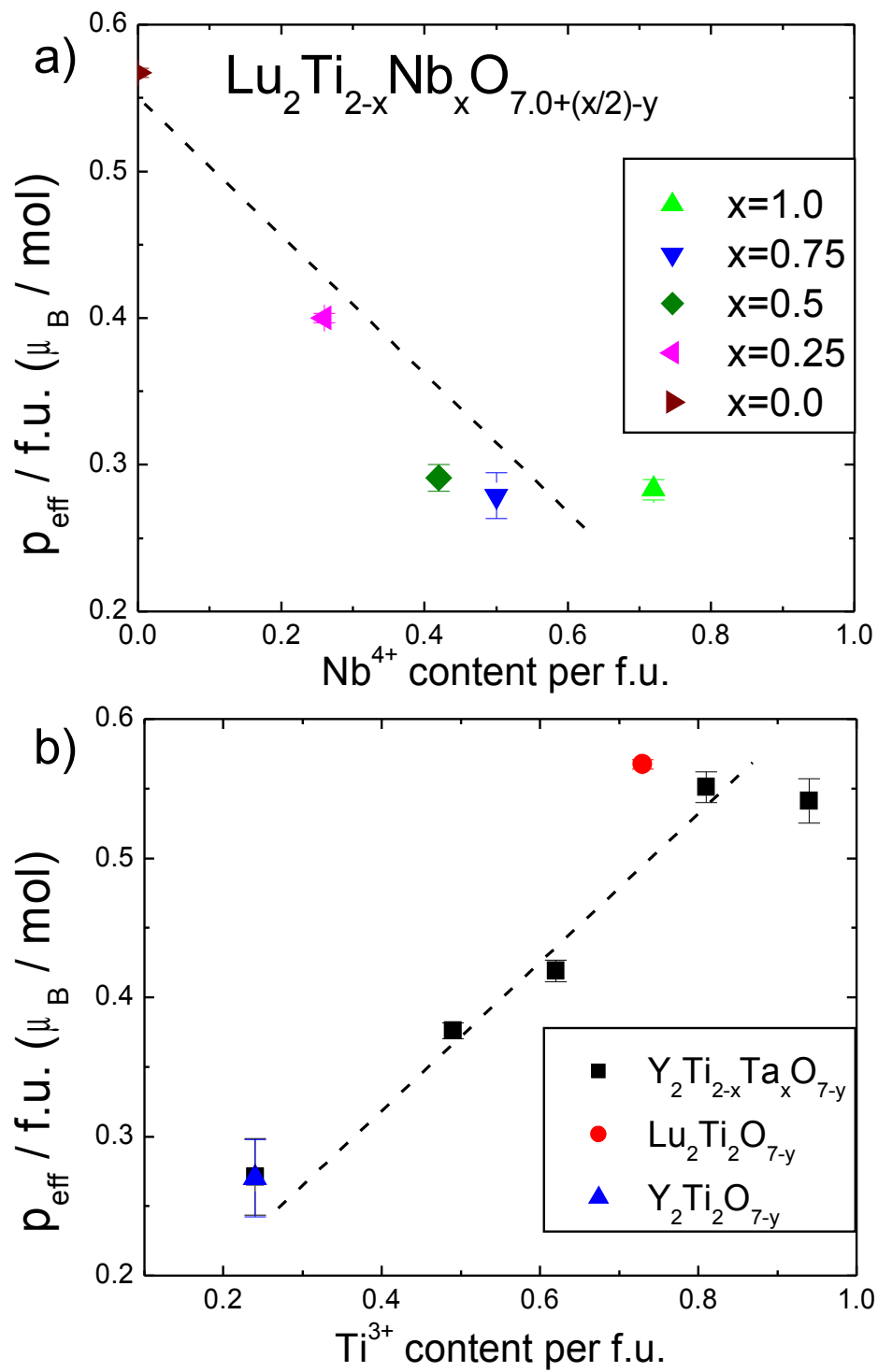

**Figure 7.**

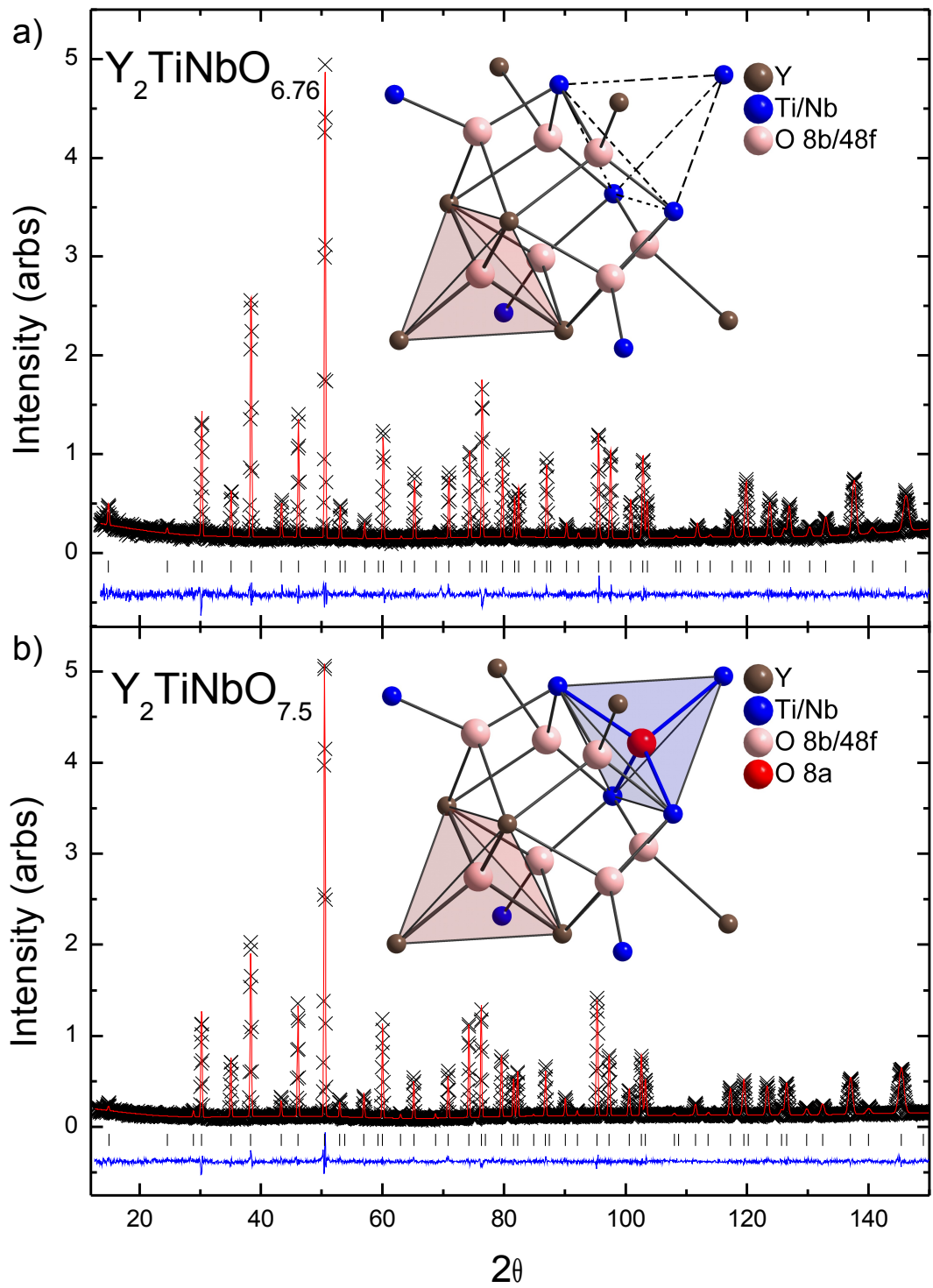